\documentclass[useAMS,usenatbib]{mn2e}
\usepackage{amsmath,fleqn,graphicx,amssymb}
\def\be{\begin{equation}} 
\def\ee{\end{equation}}

\def\bolth{\mbox{\boldmath$\theta$}}
\def\bolom{\mbox{\boldmath$\Omega$}}
\def\bolJ{\mathbf{J}}
\def\boln{\mathbf{n}}

\def\Gyr{\,\mathrm{Gyr}}
\def\kpc{\,\mathrm{kpc}}
\def\kms{\,\mathrm{km\,s}^{-1}}
\def\msun{\,{\rm M}_\odot}
\title[Disassembling the Galaxy]{Disassembling the Galaxy with angle-action coordinates}

\author[P.~J.~McMillan \& J.~J.~Binney]{
  Paul~J.~McMillan\thanks{E-mail: p.mcmillan1@physics.ox.ac.uk},
  and James~J.~Binney \\
  Rudolf Peierls Centre for Theoretical Physics, 1 Keble Road,
  Oxford, OX1 3NP, UK
}

\begin{document}
\maketitle

\begin{abstract}

Angle-action coordinates are used to study the relic of an $N$-body 
simulation of a self-gravitating
satellite galaxy that was released on a short-period orbit within the disc of
the Galaxy. Satellite stars that lie within $1.5\kpc$ of the Sun are confined
to a grid of patches in action space. As the relic phase-mixes for longer,
the patches become smaller and more numerous. These patches can be seen even
when the angle-action coordinates of an erroneous Galactic potential are
used, but using the wrong potential displaces them. Diagnostic quantities
constructed from the angle coordinates both allow the true potential to be
identified, and the relic to be dated. Hence when the full phase space 
coordinates of large numbers of
solar-neighbourhood stars are known, it should be
possible to identify members of particular relics from the distribution of
stars in an approximate action space. This would then open up the possibility
of determining the time since the relic was disrupted and gaining better 
knowledge of the Galactic potential.

The availability of angle-action coordinates for arbitrary potentials is the
key to these developments.  The paper includes a brief introduction to the
torus technique used to generate them.

\end{abstract}


\section{Introduction}\label{sec:intro}

Within the remarkably successful $\Lambda$CDM model, galaxy formation
is a hierarchical process. Large galaxies, such as the Milky Way, are
built up in mergers and the accretion of smaller building blocks
\citep[e.g.][]{WhiteRees1978,SpringelHernquist2003}. The signatures of
these processes should still be visible today in the form of
substructure such as streams in all components of the Milky Way
\citep[e.g.][]{LyndenBell1995,FreemanBH2002,Helmietal2003,Abadietal2003}.

Evidence of the wealth of substructure in the stellar halo of the Milky Way
has increased dramatically over the past 15 years, most notably with
observations of the disrupting Sagittarius dwarf galaxy
\citep{IbataGilmoreIrwin1994}, and of the streams visible in the SDSS data
\citep[e.g.][]{FieldofStreams}. Within the disc several
substructures are known, as are several mechanisms that might be responsible
for them. Stars are born in
clusters within the disc, and over a period of several Galactic rotations
these clusters evaporate and the stars become phase mixed, spreading in space
but retaining closely related orbits. This is commonly referred to as a
supercluster -- it is likely the Hyades-Pleiades supercluster formed this way
\citep{Famaeyetal2005}. Substructure can also be created by dynamical
interaction with spiral arms, or a rotating bar component; for example the
Hercules stream is thought to be associated with the bar of the Milky Way
\citep{Dehnen1999:Bar,Dehnen2000:OLR,Fux2001}.  It has been suggested that
the Arcturus group \citep{Eggen1971} is debris from a merged satellite
\citep{NavarroHelmiFreeman2004}.

Many methods exist for finding this substructure, often based on
incomplete phase space information about the stars. In the outer parts
of the halo it is possible to find substructure with only knowledge of
stellar positions on the sky, sometimes in conjunction with
photometric data \citep[e.g.][]{FieldofStreams}, or radial velocity
measurements \citep[e.g.][]{LyndenBell1995,HelmiWhite1999}. In the solar
neighbourhood, approaches that look for common proper motions have
been widely used, \citep[e.g.][]{Chereuletal1999}.

With the availability of full 6D phase space information for an increasing
number of stars in the solar neighbourhood, most notably the catalogues
resulting from the Geneva-Copenhagen and RAVE surveys
\citep{Nordstrometal2004,Steinmetz2006,Zwitter2008}, and with the prospect of
a further increase by several orders of magnitude when Gaia data become
available \citep{GAIA2001}, it is appropriate to consider methods for using
these data in full.

As discussed in \cite{Helmietal1999}, the space of integrals of motion
is a very promising one for finding substructure such as superclusters
or merger debris. Stars with a single small progenitor (star cluster or
small galaxy) will have very similar values for the integrals, which
will ensure they are tightly bunched in integral space even after phase
mixing has produced a spatial distribution that is effectively
featureless. There are additional benefits to using quantities that are not
only  integrals but adiabatic invariants, as these are
more likely to remain constant as the Galactic potential changes over
time.

Previous work has focused on spaces defined by $L_z$ (which is an
adiabatic invariant in an axisymmetric
potential), and other quantities that can be used as approximate
integrals of the motion, such as the total angular momentum
\citep[only an integral of the motion in a spherically symmetric
potential,] []{Helmietal1999}, or are not adiabatic invariants, such
as the energy \citep{HelmideZeeuw2000} or the apocentre and
pericentre of an orbit \citep{Helmietal2006}.

In this paper we demonstrate the use of angle-action coordinates to find
substructure in the solar neighbourhood.  Actions are adiabatic invariants,
and their conjugate variables, the ``angle coordinates'', increase linearly
in time.  As \cite{Tremaine99} has pointed out, these properties make them
exceptionally useful for analysing tidal streams. The difficulty of
determining the actions of stars in non-spherical potentials has, however,
severely restricted their application to date. This is the first in a series
of papers in which we show how the concept of orbital tori
\citep{McGB90,KaasB94} makes it possible to exploit the power of angle-action
coordinates for practical galactic problems.  

In Sections~\ref{sec:actang}~\&~\ref{sec:fit} we briefly introduce angle-action
coordinates, and explain how we find them for stars with known phase space
positions. In Section~\ref{sec:sim} we apply them to simulated data of a
satellite merger. We show that stars from the satellite that are observed in
the solar neighbourhood are confined to a grid of patches in action space,
even when rather a poor approximation to the Galactic potential is used. We
show further that a diagnostic defined in terms of the angle coordinates
enables the true potential to be distinguished from the false one.
Section~\ref{sec:discuss:secular} discusses minor modifications to the analysis that
are required to accommodate secular evolution of the Galactic potential over
the life of a relic.

\section{Angle-action coordinates} \label{sec:actang}

Three actions $J_i$ and three conjugate angles coordinates $\theta_i$ provide
canonical coordinates for six-dimensional phase space \citep[e.g.][\S3.5]{BT08}.
The conventional phase space coordinates $\mathbf{w}\equiv({\bf x},{\bf v})$ are
$2\pi$-periodic in the angles. The  actions
are conserved quantities for any orbit, and the angles
increase linearly with time:
 \begin{equation}
\bolth(t) = \bolth(0) + \bolom(\bolJ) t,
\end{equation}
 where the components of $\bolom$ are the orbital frequencies.  Thus, in
six-dimensional $(\bolth,\bolJ)$ space, a bound orbit moves only in the three
$\theta_i$ directions, over a surface that is topologically a
three-dimensional torus. We generally refer to this model of the orbit as
``the torus''; it is labelled by the actions $\bolJ$.

Angle-action coordinates exist for any time-independent, integrable
Hamiltonian. However, an analytic method of computing the transformations
between normal phase space coordinates and angle-action coordinates
$\mathbf{w} \leftrightarrow (\bolth,\bolJ)$ is only practical for the
Hamiltonian defined by the isochrone potential, which as limiting cases
includes the harmonic-oscillator and Kepler potentials
\citep[][\S3.5.2]{BT08}.

The Hamiltonian corresponding to a more realistic galaxy potential is
not generally integrable. However, most orbits in an axisymmetric
potential are approximately `regular' (non-chaotic), and thus admit three
approximate isolating integrals of motion. Consequently, it is
possible to find angle-action coordinates which
describe motion on these orbits over all interesting time-scales.

Methods for constructing angle-action tori for orbits in a general potential
have been in the literature for over a decade \citep{McGB90,KaasB94}, but
have been little utilised, primarily because of the technical challenges
these methods present. It is, however, possible to encapsulate these
technicalities so that users are protected from them, and once this has been
done, it is nearly as easy to construct angle-action coordinates for an orbit
in an axisymmetric potential as it is to numerically integrate the orbit with
a Runge-Kutta routine, or similar.

\subsection{The torus method}

In our current implementation we restrict ourselves to orbits in axisymmetric
potentials. Conservation of angular momentum, $J_\phi$, about the system's
symmetry axis then reduces the problem to that of motion in the $(R,z)$
meridional plane in the effective potential $\Phi_{\rm
eff}(R,z)\equiv\Phi(R,z)+J_\phi^2/2R^2$ \citep[e.g.][\S3.2]{BT08}.  $J_\phi$
is the third action.

We start with with a `toy' Hamiltonian, $H^T$, for which the relationship
$(R,z,p_R,p_z) \leftrightarrow (\theta_R^T,\theta_z^T,J_R^T,J_z^T)$ is known
analytically,\footnote{In this paper we refer to variables such as the
angles and actions in the toy Hamiltonian as $\bolth^T,\bolJ^T$, and those in
the target Hamiltonian as $\bolth,\bolJ$.  This notation differs from that of
\cite{McGB90} and \cite{KaasB94}, in which the \emph{toy} angles and actions
are referred to as $\bolth,\bolJ$, and those in the target potential as
$\bolth',\bolJ'$. We make this change in notation as our focus is primarily
on the \emph{application} of this machinery, rather than on how it works.}
namely that of a generalised effective isochrone potential
\begin{equation}
  \Phi_{\mathrm{eff}}^T(r,\vartheta)
  =\frac{-GM}{b+\sqrt{b^2 + (r-r_0)^2}} +
  \frac{L_z^2}{2\left[(r-r_0)\sin\vartheta\right]^2}, 
\end{equation}
 where $\vartheta$ is latitude (not to be confused with the dynamical angle
coordinates); $M$, $b$, $L_z$ and $r_0$ are free parameters of the toy
Hamiltonian.

As described in detail in \cite{McGB90}, the toy torus
$\bolJ^T=\hbox{constant}$ is distorted
into the ``target torus'' that approximates the orbit by the generating
function
\begin{equation}
  S(\bolth^T,\bolJ) = \bolth^T \cdot \bolJ + 2 \sum_{\boln >0}
  S_\boln(\bolJ)\sin{(\boln \cdot \bolth^T)}, 
\end{equation}
 where $\mathbf{n}$ is a two-vector with integer components and the notation
$\boln>0$ indicates that the sum runs over exactly half the plane, with the
origin excluded. The $S_\boln$ are free parameters of the generating
function.  The canonical transformation defined by this generating
function is
 \begin{equation}
  \bolJ^T=\frac{\partial
    S(\bolth^T,\bolJ) } {\partial \bolth^T}\;\; ; \;\;
  \bolth= \frac{\partial S(\bolth^T,\bolJ)
  }{\partial \bolJ}.  
\end{equation}
so
\begin{equation} \label{eq:JTinJ} \bolJ^T = \bolJ + 2
  \sum_{\boln>0}\boln S_\boln(\bolJ) \cos{(\boln \cdot \bolth^T)}
\end{equation}
\begin{equation}
  \bolth = \bolth^T + 2 \sum_{\boln>0} \frac{\partial S_\boln(\bolJ)}{\partial \bolJ} \sin{(\boln \cdot \bolth^T)} .
\end{equation}
 Since the transform is canonical for any values $S_\boln$, and the toy torus
is ``null'' in the sense that Poincar\'{e}'s integral $\int_A
d\mathbf{p}\cdot d\mathbf{q}$ vanishes for any region $A$ of the torus, the image
of the toy torus under the canonical map (the target torus) is also null.
Note that $\bolJ$ is constant on the target torus, so in general $\bolJ^T$ is
a non-trivial function of $\bolth^T$.

The values $S_\boln(\bolJ)$ (and the parameters of $H^T$) corresponding to
a given gravitational potential and $\bolJ$ are found by enforcing the
condition that the Hamiltonian $H$ is constant on the target
torus. Remarkably this condition is sufficient to ensure that the
torus corresponds to an orbit in the given galaxy potential. In practice we
enforce this condition by using the Levenberg-Marquardt algorithm
\citep[e.g.][]{Pressetal1986} to minimize the statistic $\chi^2 =
\frac{1}{N_{\rm p}}\sum(H-\overline{H})^2$, where the sum is over $N_{\rm p}$
points spread evenly in $\bolth^T$ over the target torus. The derivatives of
$H$ with respect to the $S_\boln$ and the parameters of the toy
Hamiltonian that  the Levenberg-Marquardt algorithm requires can all be found
through the chain rule. We generally refer to this process as an
``action fit''.

This minimisation determines the functional dependence
$\mathbf{w}(\bolth^T,\bolJ)$. However, it does not tell us how $\mathbf{w}$
depends on $\bolth$: the $S_\boln$ have been determined for a single value of
$\bolJ$, so $\partial S_\boln(\bolJ)/\partial \bolJ$ is still undetermined.
It is, however, possible to find \emph{approximate} values for the
frequencies, $\bolom$, by performing an orbit integration over several
periods in the target potential, starting from a phase space point on the
torus, and performing a linear fit to the resulting values of
$\theta_i^T(t)$.

To find more accurate frequencies and expressions for the angle coordinates,
we integrate small sections of the orbit, starting from a grid of points on
the torus, and use the equations \citep{KaasB94}
 \begin{equation}
  \bolth(0) + \bolom t  = \bolth^T(t) + 2\sum_{\boln>0}
  \frac{\partial S_\boln(\bolJ)}{\partial \bolJ} \sin[\boln\cdot\bolth^T(t)].
\end{equation}
 Each integration for $M$ time-steps yields $3M$ such equations in which
$\bolth(0)$, $\bolom$, and ${\partial S_\boln(\bolJ)}/{\partial \bolJ}$
are unknowns.  The equations are linear in the unknowns and for $M\gg1$ the
number of available equations increases much faster than the number of
unknowns. We truncate the sum over ${\bf n}$ to ensure that the 
number of unknowns is significantly less than the number of equations, 
and solve the equations using a least squares fit.  We
refer to this process as an ``angle fit''.

\section{Fitting the data} \label{sec:fit}

\begin{figure}
 \centerline{\includegraphics[width=.6\hsize]{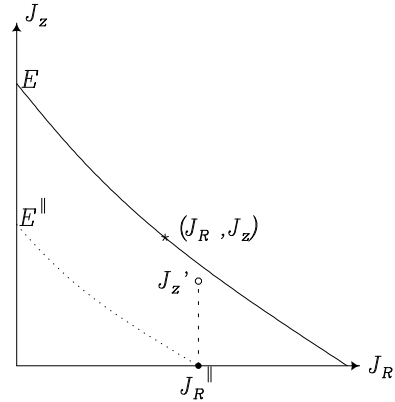}}
 \caption{
   Iterative procedure for determining $\bolJ(\mathbf{w})$. We start by
   evaluating the planar orbit with energy $E^\parallel$ (black dot).
   From its radial
   action and  vertical frequency we move to the open circle
   $(J_R^\parallel,J_z')$. Further vertical moves are used to reach the line
   on which the energy is that of the star. Then we move along that line
   to the star, where the velocities of the orbit agree with those of the star.
   \label{fig:JRJz}}
\end{figure}

The torus-fitting mechanism enables us to find $\mathbf{w}(\bolth,\bolJ)$ for
given values of $\bolJ$ rather than enabling us to determine $(\bolth,\bolJ)$
given the coordinates $\mathbf{w}$ of a star. We now explain how we go from
$\mathbf{w}$ to $(\bolth,\bolJ)$ by an iterative procedure, which is somewhat
ad-hoc, but converges quickly.  We have the numerical value $L_z$ of
$J_\phi$, so the problem is to find a point in the slice through action space
$J_\phi=L_z$. Fig.~\ref{fig:JRJz} shows this slice. Several lines of constant
$H$ are shown: the full line is the line that corresponds to the energy $E$
of the given star. The dotted line is for the ``planar energy''
$E^{\parallel}\equiv\frac{1}{2}v_{R}^2+\Phi_{\rm eff}(R,0)$ that the star
would have if it were confined to the Galactic plane. Simple one-dimensional
integrals enable us to find the action $J_R^\parallel$ and time-averaged
radius $\bar{R}$ of the orbit represented by the intersection of the dotted
line with the $J_R$ axis.  Our iterations start at this point, as 
$J_R^\parallel$ is typically a good estimate for the true value of $J_R$. 

From there we move vertically up towards the full line by an amount
$J_z'$. Bearing in mind that $\Omega_z=\partial H/\partial J_z$, we estimate
$J_z'$ from the first-order expression $E\simeq H(J_R^\parallel,J_z)\simeq
E^\parallel+\Omega_zJ_z'$.  That is, we take $J_z'=(E-E^\parallel)/\nu_{\bar{R}}$,
where we have approximated $\Omega_z$ by the vertical epicycle frequency
$\nu$ at $\bar{R}$.

We obtain the torus $(J_R^\parallel,J_z',L_z)$ and using its energy $E_2$ and
frequency $\Omega_z$ we obtain an improved approximation to $J_z$ by
incrementing $J_z'$ by $\Delta J_z'=(E-E_2)/\Omega_z$. This step is repeated
until the energy of the current torus is sufficiently close to $E$.

Once we have converged onto the line $H=E$ in Fig.~\ref{fig:JRJz}, we can
move along it with increments $\Delta\bolJ$ to $\bolJ$ that satisfy 
(to first order) 
$\bolom\cdot\Delta\bolJ=0$ -- in this process only the $R$ and $z$ components
of $\bolJ$ are changing, so $\Delta\bolJ$ is determined by $\Delta J_z$.

If the orbit does not go through the location of the star, $\Delta J_z$ is
increased when the maximum height at the star's radius is too small, and
decreased when the orbit does not reach the star's radius. Once the orbit
reaches the star, $J_z$ is adjusted until the local value of $v_z$ agrees
with the observational value. 

This procedure has  converged for all values of $\bolJ$ that we have tried,
(irrespective of, for example, whether $J_z$ is small)
and typically involves $\sim 20$ torus fits per star. With our current torus-finding
code (which we have not attempted to optimize for speed) the procedure
requires $\sim 15 s$ per star, so even now actions could be found for tens of millions of
stars in of order a week on a cluster of 1000 processors.

After fitting a torus to the observations, we obtain 
the star's angle coordinates and  more accurate frequency values by 
performing an angle fit.

\begin{figure}
  \centerline{\resizebox{\hsize}{!}{\includegraphics{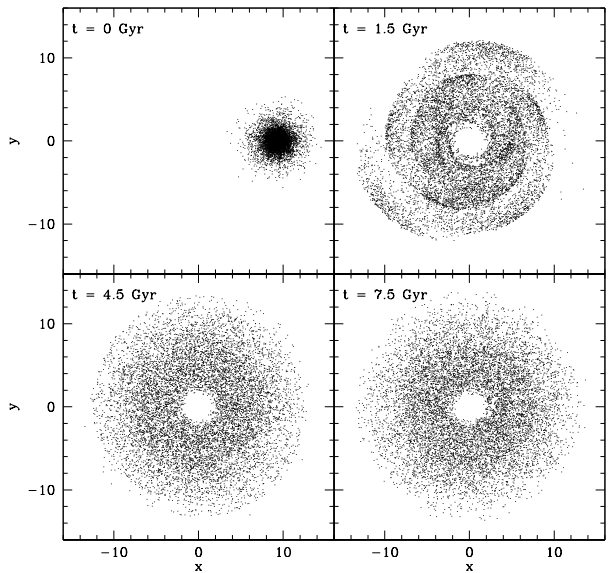}}}
  \caption{
    Particle positions in the $x$-$y$ (Galactic) plane
    initially and after $t=1.5,4.5$ 
    and $7.5\Gyr$ (as labelled).
    \label{fig:pos}}
\end{figure}

\section{Application to simulation data}\label{sec:sim}

To illustrate what can be achieved with the torus method, we use it to
examine the debris of a self-gravitating satellite that was disrupted during
an $N$-body simulation in which the satellite moved in a fixed Galactic
potential. We focus on a case that is very similar to that described in
Section~3.2 of \cite{Helmietal2006}, which was designed to reproduce the
properties of the Arcturus group.

We represented the satellite by $5\times 10^5$ particles in a King
sphere of concentration $c = \log_{10}(r_{\rm t}/r_{\rm c}) = 1.25$, core radius
$0.39\kpc$, and total mass $3.75\times 10^8\msun$. It was placed
on an orbit in the plane of the Galaxy that has apocentre at $9.3\kpc$,
pericentre at $3.1\kpc$, and angular momentum
$970\kpc\kms$. The satellite is initially placed at
apocentre and followed for $\sim 9\Gyr$.

The self-gravity of the satellite was found using $\textsc{gyrfalcon}$
\citep{Dehnen2002} and the static Galaxy potential was that of Model 2 in
\cite{DehnenBinney1998}. This model is axisymmetric, and consists of
somewhat flattened spheroids representing the halo and bulge, and three
exponential disc components to represent the gas disc and the thin and thick
stellar discs.

The particle positions in the $x$-$y$ plane are plotted in Figure~\ref{fig:pos} 
for the initial conditions and at $t=1.5,4.5$ and $7.5\Gyr$. After $1.5\Gyr$
the satellite is spread over all azimuths and particles are found over their 
entire radial range, but substructure is clearly visible. After $4.5\Gyr$
phase mixing has progressed to the extent that this structure is nearly
undetectable from the physical positions alone. 

We calculated angles, actions and frequencies for satellite particles using
both the true potential and a rather different potential -- a Miyamoto-Nagai
potential (Section~\ref{sec:MN}).

\begin{figure}
  \centerline{\resizebox{\hsize}{!}{\includegraphics[angle=270]{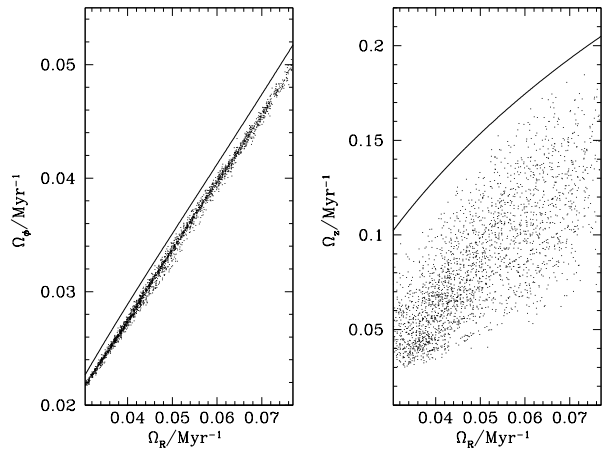}}}
\caption{ 
  $\Omega_\phi$ (left) and $\Omega_z$ (right) 
  plotted against $\Omega_R$ for a random sample of particles 
  from our simulation.
  The lines shows the the rotational frequency of a circular orbit 
  $\Omega_{\mathrm{circ}}$ (left) and vertical frequency $\nu$ (right)
  plotted against the epicycle frequency $\kappa$.
   \label{fig:omranann}
 }

\end{figure}

\begin{figure*}
  \centerline{\hfil
    \resizebox{53mm}{!}{\includegraphics{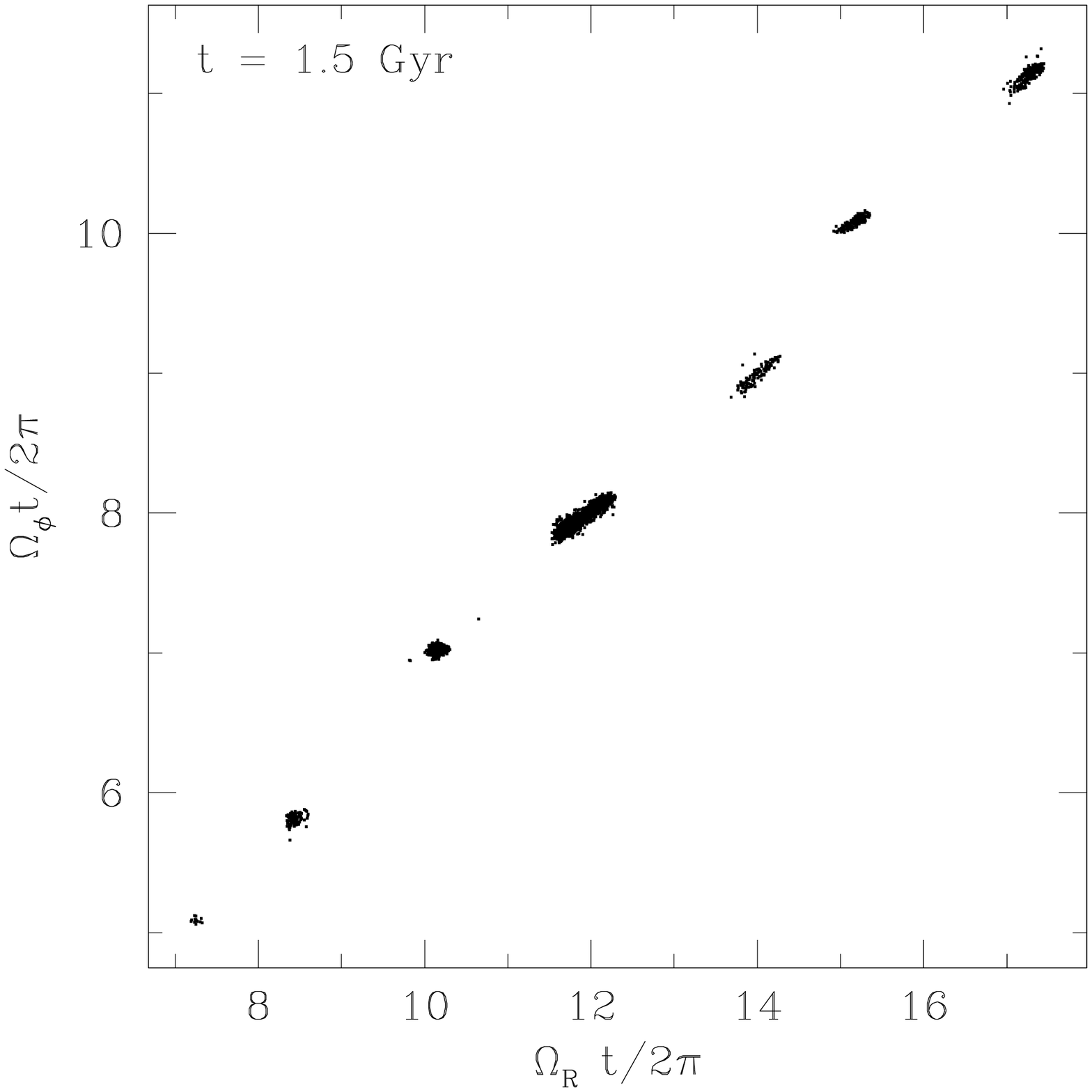}}\hspace{4mm}
    \resizebox{53mm}{!}{\includegraphics{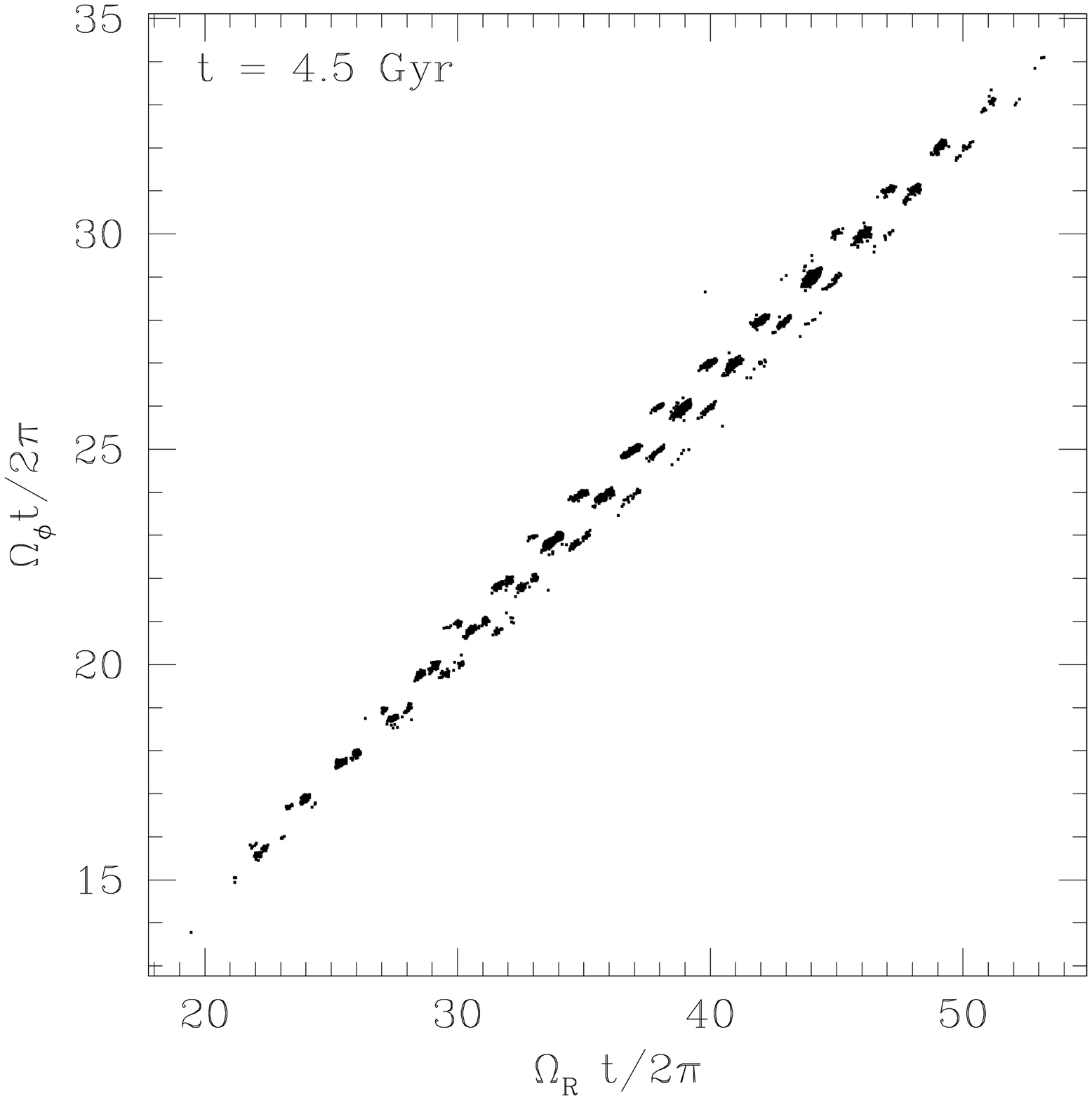}}\hspace{4mm}
    \resizebox{53mm}{!}{\includegraphics{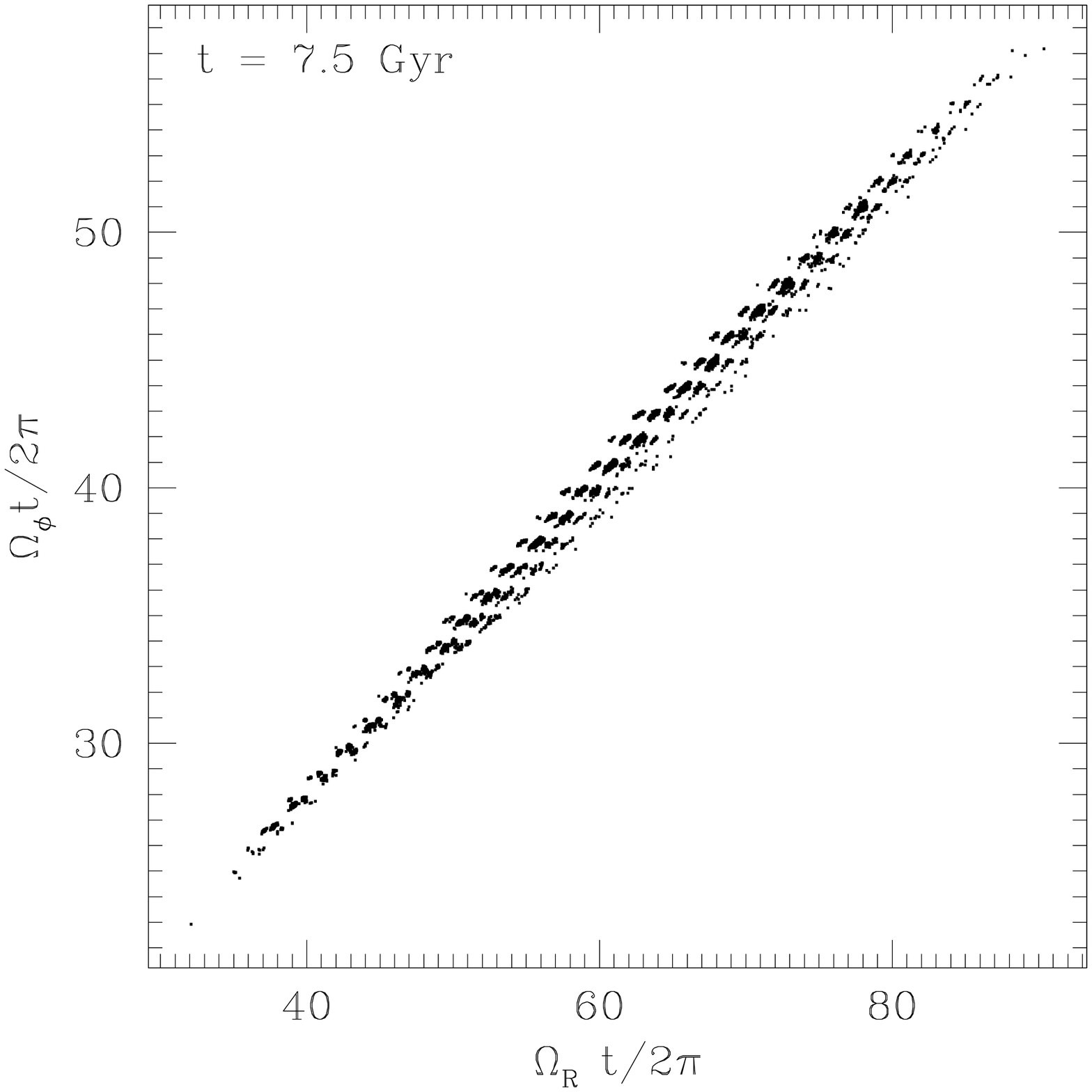}}}
  \caption{ 
    $\Omega_\phi t/2\pi$ (number of azimuthal periods) 
    plotted against $\Omega_Rt/2\pi$ (number of radial periods) for particles
    within $1.5\kpc$ of the ``solar position'' in our simulation
    after $1.5$, $4.5$ and $7.5\Gyr$ (left to right). The particles
    are separated into patches corresponding to those particles which
    have performed (approximately) an integer number of rotations about
    the Galactic centre, and are at the appropriate point in their radial
    oscillations.
   \label{fig:omsol}
 }
\end{figure*}

\subsection{The frequencies}\label{sec:freq}

Fig.~\ref{fig:omranann} shows a random sample of satellite particles at 
$t=9\Gyr$ in the $(\Omega_\phi,\Omega_R)$ plane (left) and the 
$(\Omega_z,\Omega_R)$ plane (right) when the true potential is used. 
The full lines show the relationship between the epicycle frequency 
$\kappa\simeq\Omega_R$ and the circular frequency 
$\Omega_{\mathrm{circ}}\simeq\Omega_\phi$
(left) or vertical frequency $\nu\sim\Omega_z$ (right).
These demonstrate that the strong correlation between $\Omega_R$ and 
$\Omega_\phi$
arises because each frequency depends strongly on energy and only much more
weakly on either eccentricity or inclination to the plane.  The relationship
between $\Omega_z$ and the other frequencies is much broader, reflecting the
strongly anharmonic nature of vertical oscillations, which cause the vertical
frequency to depend strongly on vertical amplitude. Because the 
simulated satellite is on an orbit in the Galactic plane, 
$\Omega_z$ is of little further interest in this
study (though it will be in other cases).

Consider now the frequencies of particles that lie in a given volume around
the Sun, as survey stars usually do.  We place the Sun $8\kpc$ from the
Galactic centre along the line to the satellite's initial location and select
particles that lie within $1.5\kpc$ of the Sun.  Fig.~\ref{fig:omsol} shows
these stars at $t=1.5,4.5$ and $7.5\Gyr$ in the plane spanned by $\Omega_\phi
t/2\pi$ and $\Omega_Rt/2\pi$ -- the number of rotations about the galactic
centre and the number of radial periods, respectively. Now we see a clear
substructure within the plot. Stars are found in patches at (close to)
integer intervals in $\Omega_\phi t/2\pi$, and regularly in $\Omega_Rt/2\pi$.

The reason for this clumping is simple: to be in the solar neighbourhood at
time $t$, having also all been near each other at an earlier point in time
(when part of a single satellite), these particles must all have moved a
certain amount in $\phi$, plus or minus an integer number of complete
rotations about the Galactic Centre. This is, in essence, a selection effect
caused by taking a window of finite size. Even after $7.5\Gyr$, at which
point phase mixing has rendered the spatial distribution essentially
featureless (Figure~\ref{fig:pos}), there is manifest clumping in the
$\Omega$ plot. In the case of $\Omega_Rt/2\pi$ the patches occur more
frequently than at integer intervals because the orbits cross the radial
range twice per radial period.  The non-zero size of the patches reflects the
non-zero size of the window, non-zero initial velocities of the particles
relative to the satellite motion and the non-negligible mass of the
satellite, which causes orbits to deviate from orbits in the Galactic
potential at early times. There is also a small spread due to any errors in 
the value of $\Omega$ found -- this is clearly a small effect as the
patches are still distinct after $7.5\Gyr$. The
spread in $\Omega_z$ among stars in these samples is no narrower than that of
all the satellite's stars, and not separated into patches, both because our
window constrains $z$ only weakly, and because the initial values of
$\theta_z$ range from zero to $2\pi$ as a result of the satellite starting
from the plane with $v_z=0$.

The number of patches in $\bolom$-space increases approximately as $t^2$, since
the number of integers that lie in the full range of $\Omega_Rt/2\pi$ or 
$\Omega_\phi t/2\pi$ is proportional to $t$ (since the range of $\bolom$ 
doesn't change). The size of individual patches is, to a first approximation,
determined by the size of the window from which particles are chosen: if an 
orbit lies within the window over a range $\Delta\theta_R$ (ignoring for 
simplicity the dependence on $\theta_z,\,\theta_\phi$), then in 
$\bolom$-space (as opposed to $\bolom t/2\pi$-space) each 
individual patch will have width $\Delta\theta_R/t$. 
Therefore the size of each patch in $\bolom$-space 
is proportional to $t^{-2}$, so the
total area of the patches is approximately time-independent. The
other effects mentioned above that cause the patches to have finite size
have a similar effect on the size of the patches with the exception of the error in 
measurements of $\Omega$. Some of the patches are restricted in size 
because they meet the edge of the envelope of available $\bolom$-values (the 
values found in the satellite as a whole -- Figure~\ref{fig:omranann}, left). 
This is probably most obvious in the left panel of Figure~\ref{fig:omsol},
in which one patch (at $\Omega_\phi t/2\pi = 8$) is far larger than the
others because it nowhere touches the envelope.

\begin{figure}
    \centerline{\resizebox{.8\hsize}{!}{\includegraphics{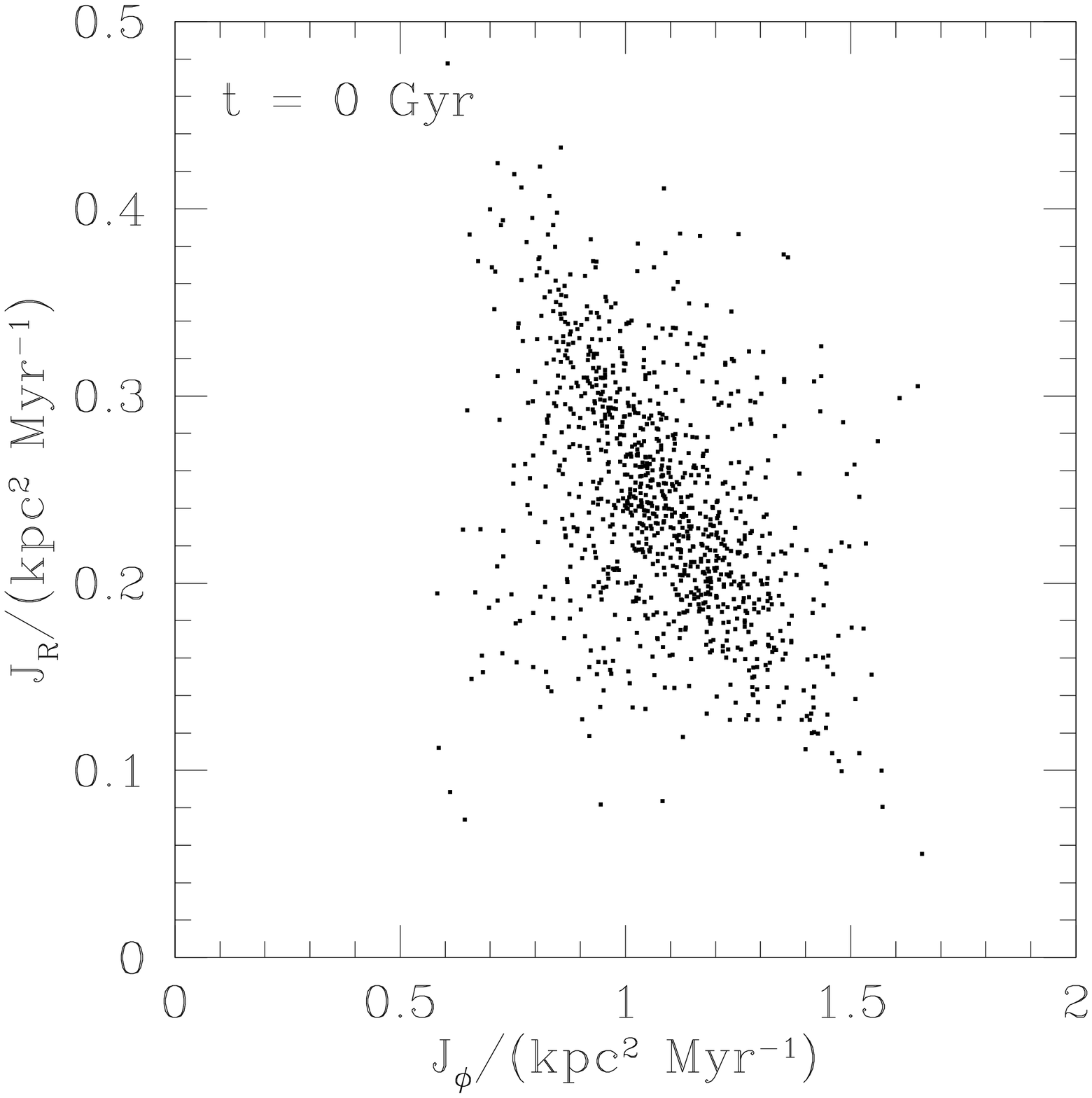}}}
    \centerline{\resizebox{.8\hsize}{!}{\includegraphics{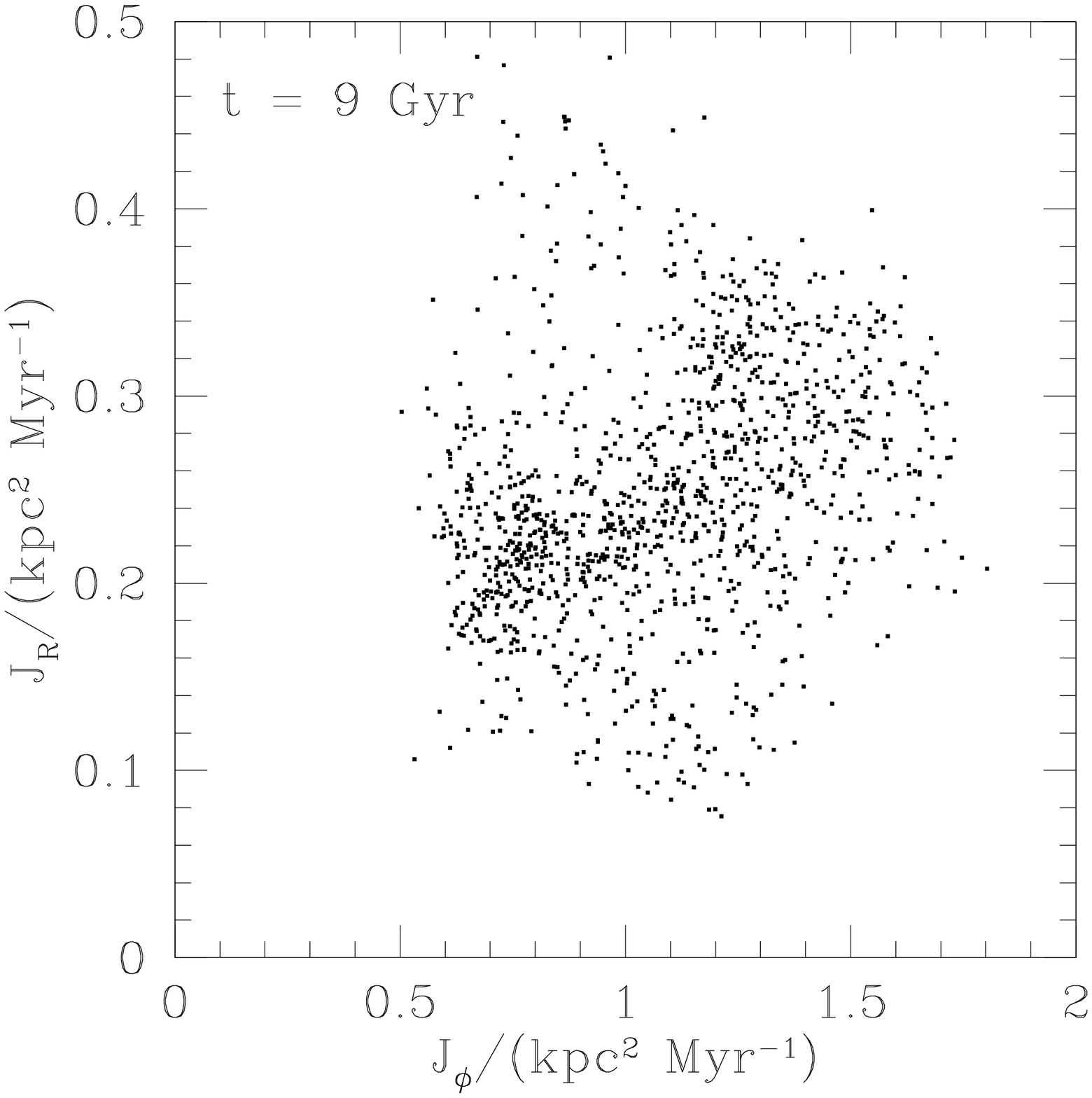}}}
  \caption{ 
    $J_R$ plotted against $J_\phi$ for a random sample of particles at the
    beginning of the simulation (top) and after $9\Gyr$ of evolution (bottom).
   \label{fig:randJ}
 }
\end{figure}

\subsection{The actions}

\begin{figure*}
  \centerline{\hfil
    \resizebox{54mm}{!}{\includegraphics{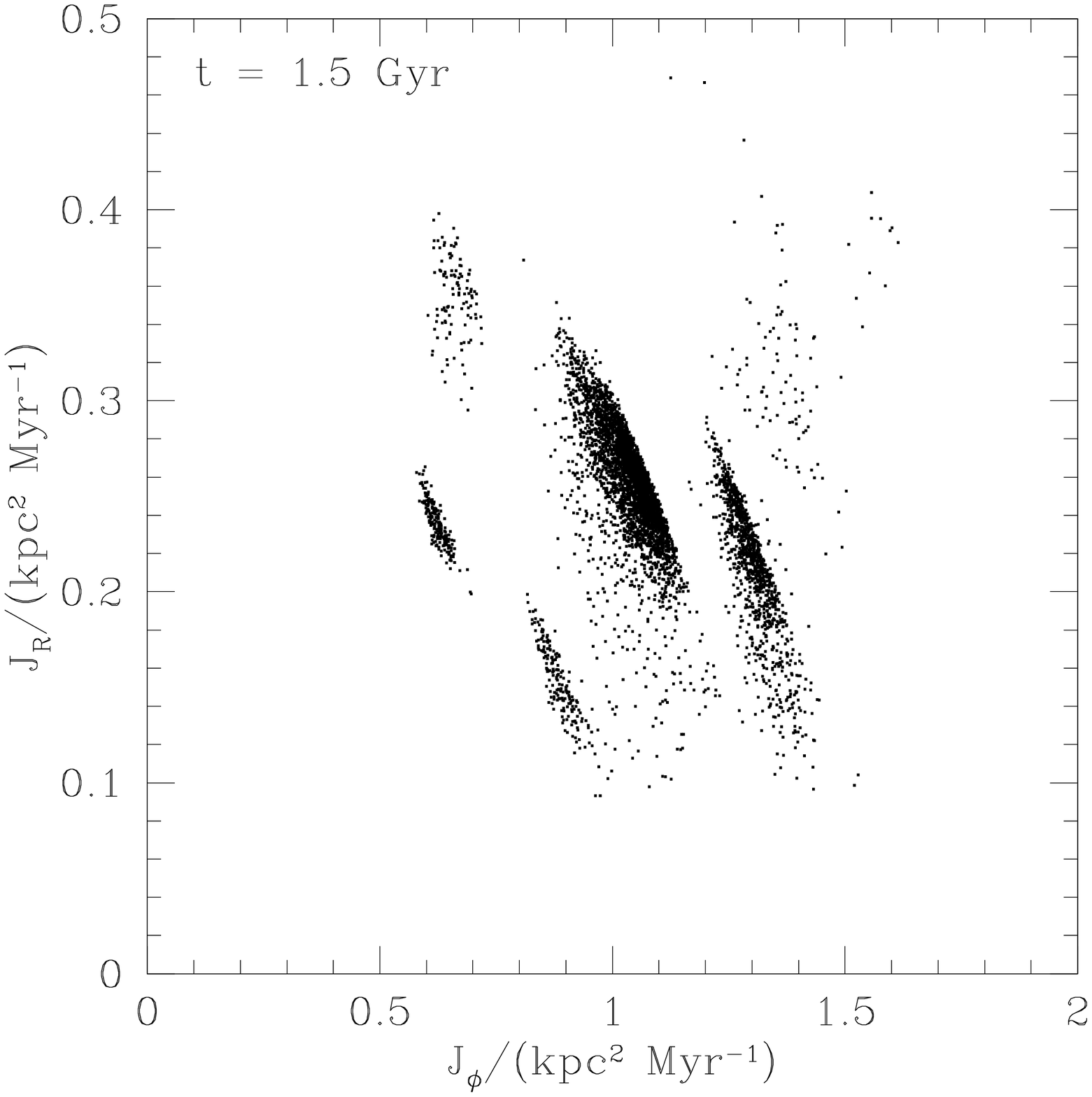}}
    \resizebox{54mm}{!}{\includegraphics{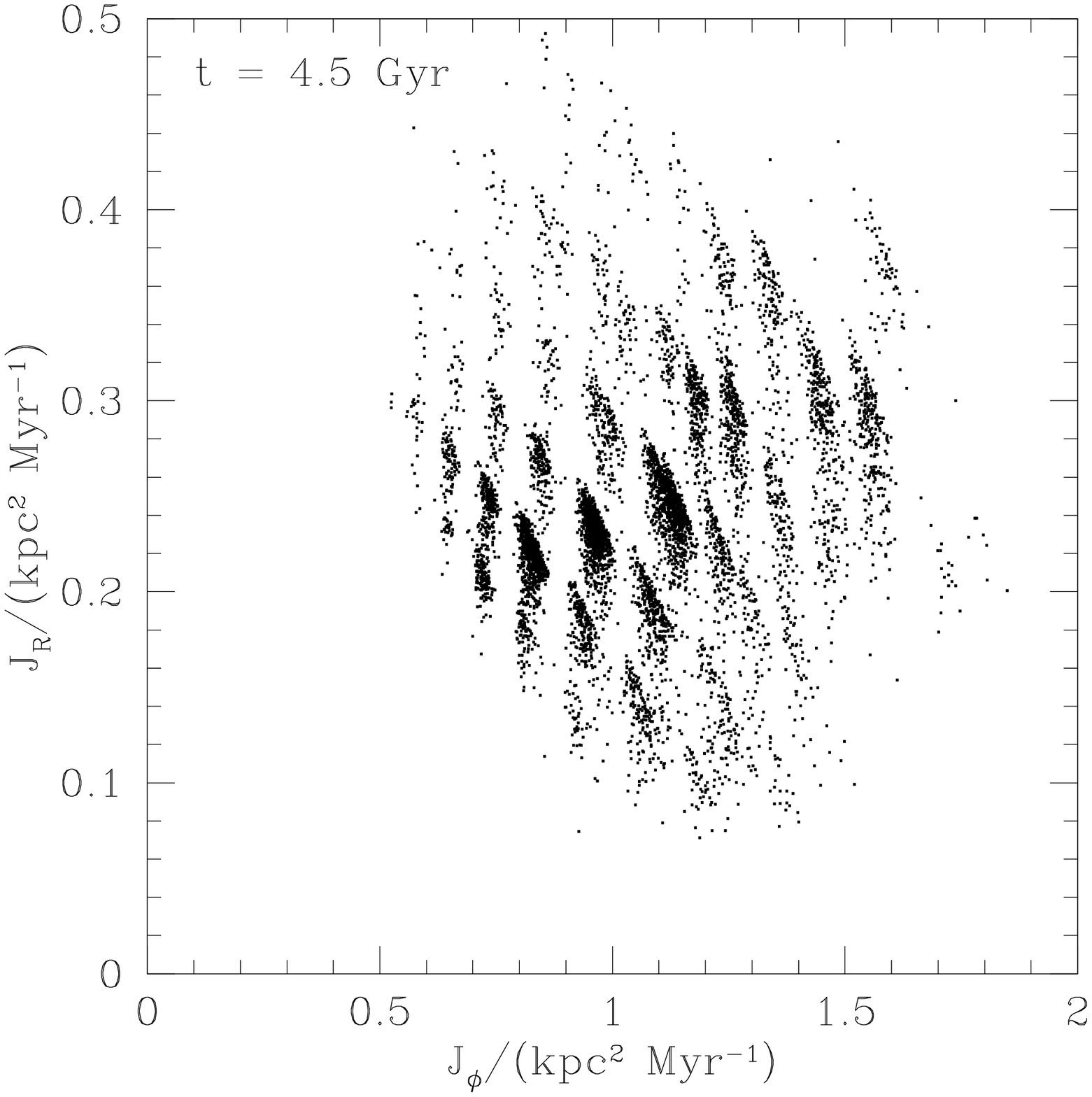}}
    \resizebox{54mm}{!}{\includegraphics{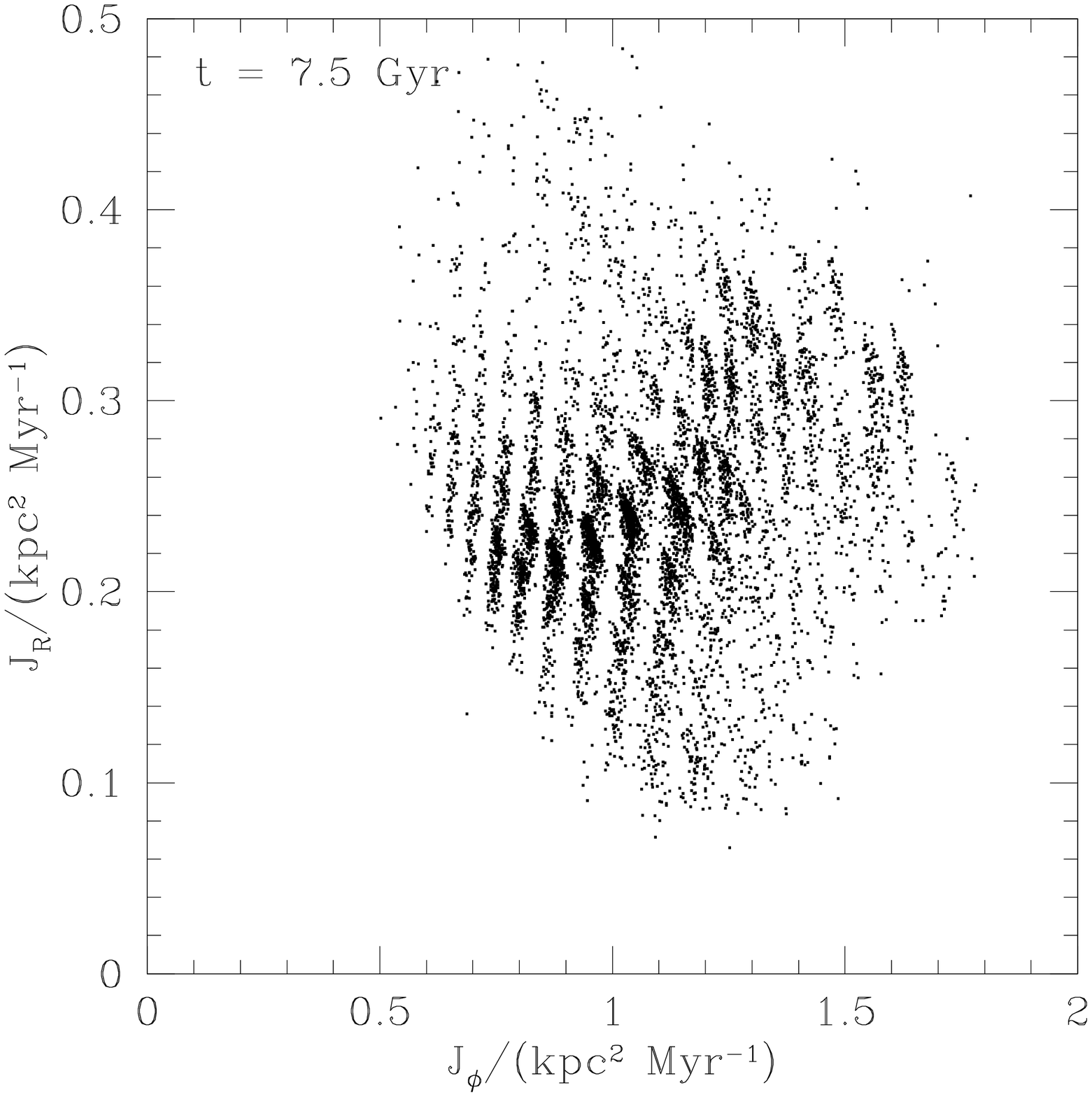}}}
  \caption{ 
    $J_R$ plotted against $J_\phi$ for particles
    within $1.5\kpc$ of the ``solar position'' in our simulation
    after $1.5$, $4.5$ and $7.5\Gyr$ (left to right). As in 
    Figure~\ref{fig:omsol}, the particles are divided into patches which
    increase in number and decrease in size as $t$ increase.
   \label{fig:Jsol}
 }
\end{figure*}

\cite{Helmietal2006} sought to identify substructure in the disc by
calculating the locations of stars in ``APL'' space, which is the space
spanned by apocentre, pericentre and $J_\phi$ (which they refer to as $L_z$). 
Apo- and pericentre can be
considered to be integrals analogous to actions, so APL space is a mapping of
action space. Hence it is of interest to examine the distribution of the
satellite's stars in action space for comparison with the results of
\cite{Helmietal2006} although we shall find it less interesting than the
frequency and angle spaces.

Fig.~\ref{fig:randJ} is a plot of $J_R$ against $J_\phi$ for a random set of
particles (reflecting the satellite as a whole) at the beginning of the
simulation (top) and at $t=9\Gyr$ (bottom). The particles remain in the
same general area of the $(J_\phi,J_R)$ plane, but at early times the actions are not
constant because the satellite is self-gravitating and the strong negative
correlation between $J_R$ and $J_\phi$ seen in the initial conditions is
replaced by a (rather weaker) positive correlation.\footnote{When the
satellite's self-gravity is turned off, the actions prove to be constant as
expected.}

The strong negative correlation in the initial conditions arises because
initially the satellite is at apocentre. A particle that moves relative to
the centre of the satellite in the opposite direction to the satellite's
rotation about the Galactic centre has less angular momentum than one that
moves in the opposite direction, and -- in the absence of the satellite's
self-gravity -- would be on a more eccentric orbit, and thus have a higher
value of $J_R$.

The weaker correlation between $J_R$ and $J_\phi$ seen at $9\Gyr$ arises because 
the actions of a particle become constant  when the particle is stripped from
the satellite and starts to feel the latter's gravity only weakly. This occurs at
pericentre, when the effect of combining motion within the satellite with the
motion of the satellite is precisely opposite of what it is at apocentre. The
extent of the correlation between $J_R$ and $J_\phi$ at $9\Gyr$ is comparable
to that shown in Fig.~5 of \cite{Helmietal2006} for the locations in APL
space of stars that lie within $5\kpc$ of the Sun.

Figure~\ref{fig:Jsol} is a plot of $J_R$ against $J_\phi$ for particles that
lie within $1.5\kpc$ of the Sun at $t=1.5,4.5$ and $7.5\Gyr$ (left to right).
The actions of the particles found in the solar neighbourhood cover almost
the entire range in $\bolJ$ found in the satellite as a whole, but are
separated into distinct patches that decrease in size and increase in number
over time.  The number of patches increases slightly faster than the area of
each patch decreases with the result that by $7.5\Gyr$ the patches are
starting to merge into bands.  Fig.~5 of \cite{Helmietal2006} shows such a
series of bands in APL space for stars that lie within $1.5\kpc$ of the Sun.

\begin{figure}
  \centerline{\resizebox{\hsize}{!}{\includegraphics{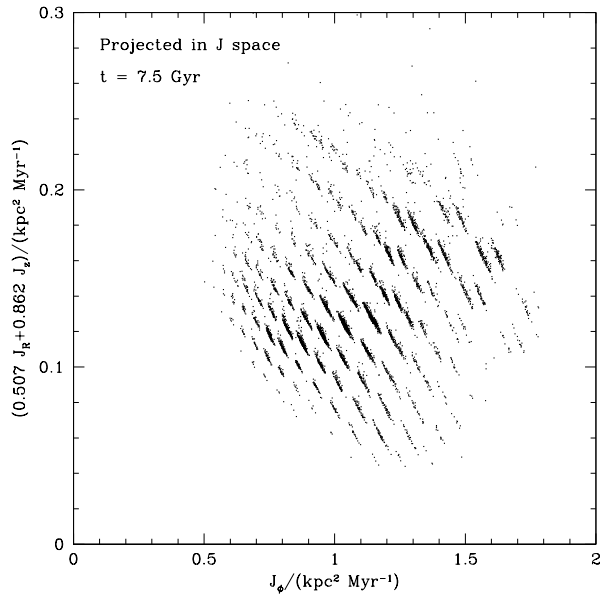}}}
  \caption{
    A projection in $\bolJ$ space of the actions of particles
    within $1.5\kpc$ of the ``solar position'' in the simulation
    after $7.5\Gyr$ (the same particles plotted in 
    Figure~\ref{fig:Jsol}, right). In this projection it is clearer that
    the stars are separated into individual clumps in $\bolJ$ space.
    \label{fig:angleJ}}
\end{figure}

Since $\bolom$ is a smooth function of $\bolJ$, our study of the distribution
of particles in frequency space explains their distribution in action space:
the ``allowed'' values of $\bolJ$ correspond to ``allowed'' values of
$\bolom$, which are confined to patches. While the constraints on $\bolom$ do
not involve $\Omega_z$, the constraints on $\bolJ$ \emph{do} depend on $J_z$,
because both $\Omega_R$ and $\Omega_\phi$ depend on $J_z$. Consequently,
the positions of stars in $\bolJ$-space form a relatively regular lattice,
but the principal directions of that lattice are not
parallel to the $J_i$ axes. Therefore, the tendency of
patches to run together in the extreme right-hand panel of
Fig.~\ref{fig:Jsol} can be eliminated by plotting a different projection of
action space. For example, Figure~\ref{fig:angleJ} is a plot of
$0.507J_R+0.862J_z$ against $J_\phi$ -- a projection chosen by eye from a 3D
visualisation of the distribution in $\bolJ$-space -- and in this plot 
the patches are all distinct. In general, the patches will be most cleanly
separated when the lattice is projected  along one of it principal directions,
since then
points with the same $\Omega_R$ and $\Omega_\phi$ but differing in $\Omega_z$
are projected on top of one another. The optimum projection
depends both on the potential and on the region of $\bolJ$-space occupied by
the stars, but it  can be straightforwardly identified  for any set of data
because $\bolom$ is found at the same time as $\bolJ$.

\begin{figure}
      \resizebox{82mm}{!}{\includegraphics{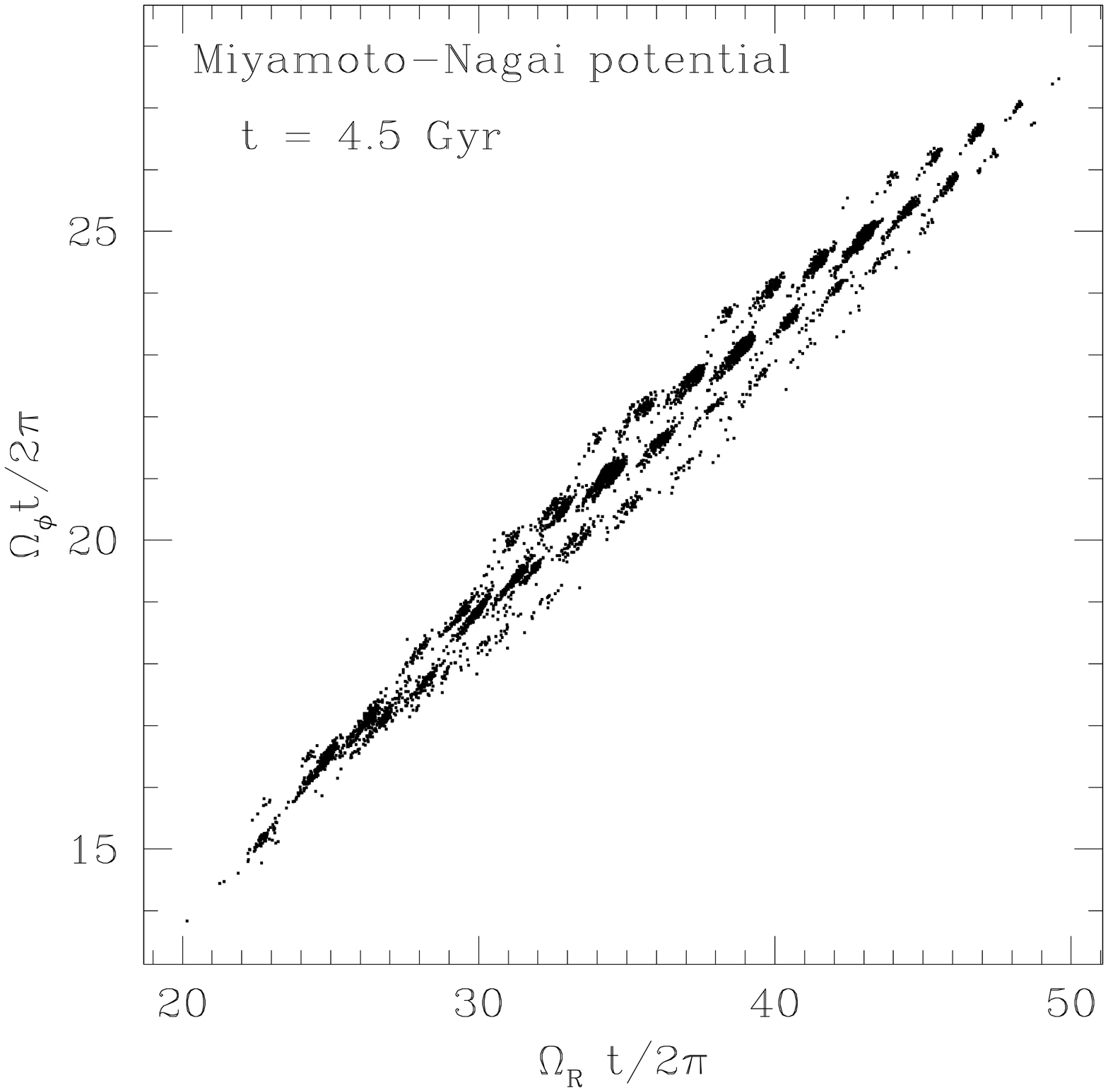}}\hspace{1mm}
    \resizebox{82mm}{!}{\includegraphics{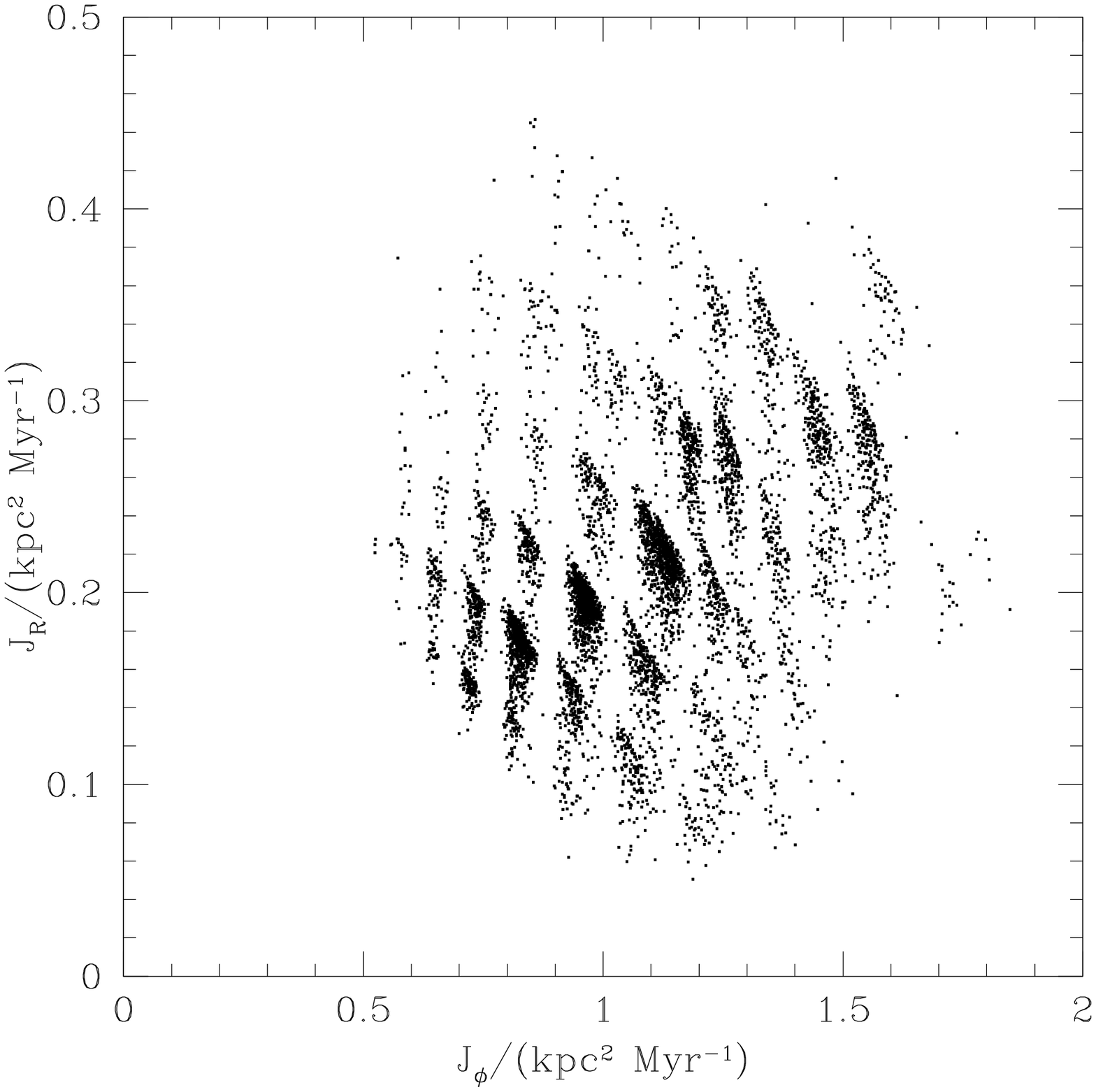}}
  \caption{ 
    $\Omega_\phi t/2\pi$ (number of azimuthal periods) 
    plotted against $\Omega_Rt/2\pi$ (top), and $J_R$ against $J_\phi$
    (bottom) for particles found within $1.5\kpc$ of the ``solar position'' 
    in our simulation after $4.5\Gyr$ as determined in the Miyamoto-Nagai 
    potential described in Section~\ref{sec:MN} (for comparison see the
    middle panels of Figures~\ref{fig:omsol}~\&~\ref{fig:Jsol}).
   \label{fig:MN}
 }
\end{figure}

\subsection{Working with incomplete knowledge} \label{sec:MN}

In reality we do not know the Galaxy's potential a priori. In this subsection
we show that satellite particles can be identified using even a poor
approximation to the potential, and then the true potential identified from
structure within the sample of satellite particles.

We repeated the above analysis using 
orbital tori  in the
Hamiltonian for a Miyamoto-Nagai potential
 \begin{equation}\label{eq:MN}
  \Phi_{\rm MN}(R,z) = -\frac{GM}{\sqrt{R^2 + (a+\sqrt{z^2+b^2})^2}},
\end{equation} 
 with mass $M=1.8\times10^{11}\msun$, scale length $a=6\kpc$ and
scale height $b=0.3\kpc$. This is a crude approximation to the true
potential and one expects to be able to start from a better approximation to
the Galaxy's potential.  It is chosen such that the circular speed at the
Solar radius is approximately the same as in the true potential, and the
scale height is similar to that of the true disc. We chose a scale length
that is much greater than that of the true thin disc (which dominates the
forces in the solar neighbourhood) as the Miyamoto-Nagai potential falls off
quickly with radius, and we want to avoid any risk of having particles at or
above the escape speed (at least one action diverges as a particle's speed
tends to the escape speed).

Fig.~\ref{fig:MN} shows plots of $\Omega_R$ against $\Omega_\phi$ (top) and
$J_R$ against $J_\phi$ (bottom). While the clear separation of particles into
clumps in both $\bolom$ and $\bolJ$ seen in Figure~\ref{fig:Jsol} is somewhat
smeared by using the wrong potential, it is not completely lost. Therefore,
even when the true potential is unknown, these plots enable us to identify
substructures.

We can go further, and use the displacement of patches in $\bolom$ space by
an erroneous potential to identify the true potential.  Specifically, at time
$t$ the angle coordinates of the $\alpha$th particle $\bolth_\alpha $ satisfy
 \begin{equation}\label{eq:thetalapse}
\bolom_\alpha (t-t_0)-(\bolth_\alpha -\bolth_{\alpha,0})  =  2\pi\mathbf{m}_\alpha ,
\end{equation}
 where the particle was at $\bolth_{\alpha,0}$ at time $t_0$; the (integer)
components of $\mathbf{m}_\alpha$ give the number of oscillations that the
particle has made in $R, z$ and $\phi$.

\begin{figure*}
  \centerline{\hfil
    \resizebox{\hsize}{!}{\includegraphics[angle=270]{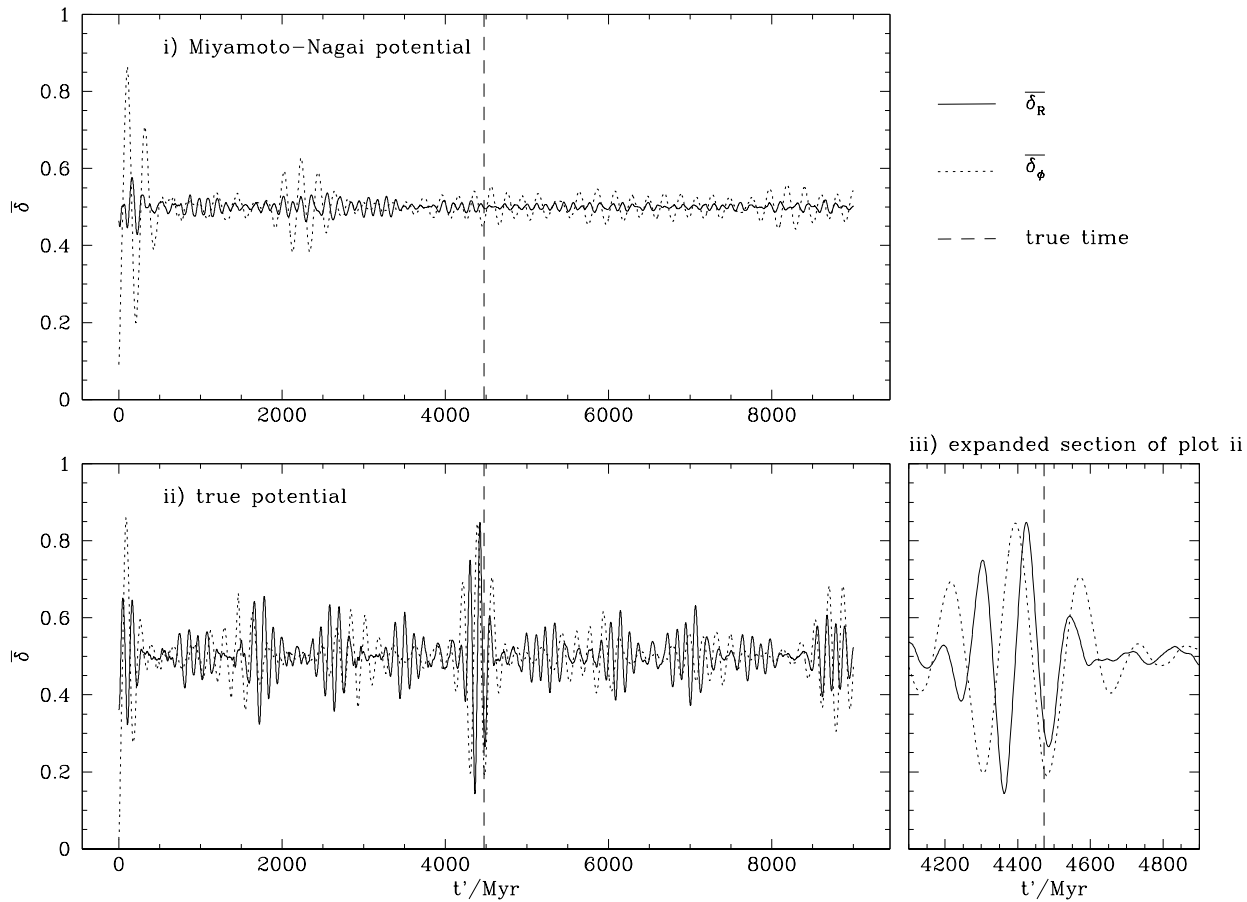}}}
  \caption{ 
    $\overline{\delta_R}$ (solid line), and $\overline{\delta_\phi}$ 
    (dotted line) plotted against $t'$ (see 
    Equation~\ref{eq:delr}); the dashed line indicates the
    known true value of t. The lower panels (ii \& iii)
    show $\overline{\delta_R}$ and $\overline{\delta_\phi}$ determined using
    values of $\bolom$ and $\bolth$ found in the same potential that the 
    orbit integration was carried out in -- the lower-right panel (iii) 
    being a magnified section of the lower left hand panel (ii), focused
    around the true value of t. The upper panel (i) shows 
    $\overline{\delta_R}$ and $\overline{\delta_\phi}$ determined using
    values of $\bolom$ and $\bolth$ found in an incorrect potential
    (Equation~\ref{eq:MN}). There are strong minima in both 
    $\overline{\delta_R}$ and $\overline{\delta_\phi}$ around the true
    value of t when the true potential is used, whereas when an 
    incorrect potential is used none is seen. 
   \label{fig:stat}
 }
\end{figure*}

There is negligible clumping in $\theta_z$ because the satellite's orbit
initially lay in the plane, but at some time $t_0$, before the satellite was
stripped, individual values of $\theta_{R,\alpha,0}$ were tightly correlated,
as were values of $\theta_{\phi,\alpha,0}$. In fact, in
our simulation at $t = 0$, the satellite was centred at
$\theta_{\phi,0} = 0$ and at apocentre, where $\theta_R \simeq \pi$ for all
orbits.  Therefore, we define the statistical measures 
\begin{eqnarray} \label{eq:delr}
\delta_{R,\alpha} &=& \left |
  \frac{\Omega_{R,\alpha}t'-(\theta_{R,\alpha}- \theta_{R,0}) -  2\pi m_{R,\alpha}}{\pi} \right
  |\nonumber\\
\delta_{\phi,\alpha} &=& \left | 
  \frac{\Omega_{\phi,\alpha}t'-(\theta_{\phi,\alpha}- \theta_{\phi,0}) -  2\pi m_{\phi,\alpha}}{\pi} \right |,
\end{eqnarray}
 where the integers $m_{\phi,\alpha}$ and $m_{R,\alpha}$ are chosen such that
$\delta_{R,\alpha}$ and $\delta_{\phi,\alpha}$ are minimised; $\theta_{R,0}=\pi$ 
and $\theta_{\phi,0}=0$ (in this case) and $t'$ is a
free parameter.  In these expressions the numerators vary between $\pm\pi$,
so when the values of $\Omega_\alpha t'$ are randomly distributed,
$\bar{\delta}_i\simeq0.5$, while when $\Omega_it'$ has a well defined phase
relative to the rest of the numerator, $\bar{\delta}_i$ can approach either
zero or 1, depending on whether the two halves of the numerator are in or out
of phase with each other.

At $t=4.5\Gyr$ in our simulation $\overline{\delta}_R$ and
$\overline{\delta}_\phi$ were evaluated by summing over particles within
$1.5\kpc$ of the Sun.  Fig.~\ref{fig:stat} shows the resulting plots of
$\overline{\delta}_R$ and $\overline{\delta}_\phi$ as functions of $t'$ with
$\bolom_\alpha $ and $\bolth_\alpha $ determined in the Miyamoto-Nagai
potential (upper panel) and in the true potential (lower panel). The upper
panel shows that when the dynamical variables are evaluated in an erroneous
potential, the $\bar{\delta}_i$ scarcely move from their mean values after
the first few megayears, and when one of them does move downwards, the other
does not. By contrast in the lower panel for the true potential both
$\bar{\delta}_i$ display sustained beats as the relative phases of the two
terms that make up the numerators in equations (\ref{eq:delr}) have stable
relative phases. In the neighbourhood of the true disruption time of the relic the
beating swells in amplitude and, as the exploded view on the right of the
figure shows, both move down together just $10\,$Myr past the relic's true
age.

When applying the test involving the $\bar{\delta}_i$ to real observational
data, it will be necessary to search for beats over both $t'$ and the mean
of the initial azimuthal phases $\theta_{\phi,\alpha}$. Hence,
Fig.~\ref{fig:stat} oversimplifies the problem because it shows a
one-dimensional search rather than a two-dimensional one. Moreover, if the
satellite was initially on an inclined orbit rather than one in the plane,
the vertical angles $\theta_z$ would be involved, so coincident beats would
be required in three angle-dependent variables $\bar{\delta}_i$ rather than
two. Consequently, a smaller fraction of a relic's stars would satisfy this
condition at a given time, making it harder to identify a relic against
background noise. The upside of the involvement of $\theta_z$ is
that it would give us the opportunity to constrain the vertical structure of
the potential; a relic of a satellite that started from $J_z\simeq0$
only probes the Galactic rotation curve. 

Real data will also have background stars that did not come from the
satellite. Naturally, the Poisson noise in the distribution of background
stars makes it harder to identify overdensities in $\bolJ$-space, but once
the overdensities associated with a remnant have been identified, background
stars have negligible impact on the ability of the $\bar{\delta}_i$ to
determine the disruption time: any star in the overdensity is in the same
part of $\bolom$-space as the remnant's stars, so the contributions to
equation~(\ref{eq:delr}) from background and remnant stars will differ only
in the relevant values of $\bolth$. These differences will inevitably be
small because we are dealing with a small survey volume. For example, if
$20\%$ of the stars identified as being part of the remnant are actually
background stars, and are -- on average -- displaced from the position on the
orbit where a typical remnant member with the same $\bolom$ would be by
$\Delta\theta_i=0.2$ (equivalent to $\sim 1.5\kpc$ in the $\phi$-direction),
this would only make a difference of $~0.01$ to $\bar{\delta}_i$.

Before the satellite is completely disrupted, it is affected by dynamical
friction, which will cause spreading in velocity space. However this is a
significantly smaller effect than that due to the self gravity and initial
velocity dispersion in the satellite for an object of this mass -- the
Chandrasekhar dynamical friction formula \citep[e.g.][\S8.1]{BT08} suggests
that a satellite of this mass should be decelerated by $\sim
10\,\mathrm{km}\,\mathrm{s}^{-1}\Gyr^{-1}$, and this decreases in direct
proportion to the satellite mass as it is stripped. Hence including dynamical
friction would shift individual velocities by significantly less than their
intrinsic scatter, namely the internal velocity distribution of the satellite
($18.6\,\mathrm{km}\,\mathrm{s}^{-1}$). 

These matters, and any others that arise, 
will be investigated further, but Fig.~\ref{fig:stat} clearly 
conveys the essential idea and gives a tantalising taste of 
the diagnostic potential of angle variables. 

\section{Secular evolution}\label{sec:discuss:secular}

Stars have continued to form at a significant rate throughout the lifetime of
the Galaxy's thin disc, and it must be presumed that the disc's mass has
increased significantly over the last $5\Gyr$. Since actions are adiabatic
invariants, such secular evolution of the Galactic potential does not affect
the distribution of relic stars in the $J_\phi-J_R$ plane
(Fig.~\ref{fig:randJ}). Secular evolution causes the frequencies at fixed
$\bolJ$ to become explicit functions of time, so we should write
$\bolom(\bolJ,t)$, and the increment in $\theta_i$ over time $t$ changes from
$\Delta\theta_i=\bolom_it$ to $\Delta\theta_i=\int_0^tdt\,\bolom_i$. The conditions
for a relic star to be in the solar neighbourhood are given values of
$\Delta\theta_i\mod2\pi$ for $i=R,\phi$.  Since with secular evolution
$\Delta\theta_i$ remains a continuous function of $\bolJ$, these conditions
continue to be satisfied only in a grid of patches in action space;
secular evolution shifts the patches, but does not blur them. Hence the
diagnostic power of plots like Fig.~\ref{fig:Jsol} is unaffected by secular
evolution. 

In the presence of secular evolution it becomes necessary in equation
(\ref{eq:thetalapse}) to replace $\bolom_\alpha(t-t_0)$ by
$\int_{t_0}^tdt\,\bolom_\alpha$. To evaluate the required integrals, one must
adopt a model of the history of the potential, which determines the time
dependence of $\bolom$.  The required model is self-evident if secular
evolution is confined to growth in the disc's mass at a known rate.
Uncertainties in this rate will make it harder to locate beats in
Fig.~\ref{fig:stat}.

Of course the key to calculating the secular evolution of stellar systems, be
they globular clusters or galaxies, is to express their distribution
functions in terms of actions \citep[e.g.][]{SellwoodMcgaugh05}, so the
availability of orbital tori for arbitrary potentials opens up new horizons
in this area.

\section{Summary} \label{sec:discuss:summary}

A major hope of ``near-field cosmology'' is to identify within the Galaxy
groups of stars that were accreted together \citep{FreemanBH2002}. We have
demonstrated the power of angle-action coordinates for doing this by studying
the debris of a self-gravitating satellite of mass $3.75\times10^8\msun$,
released within the plane of a realistic Galactic potential on an orbit with
apocentre at $9\kpc$. On this short-period orbit the satellite's stars become
well phase mixed within a couple of gigayears, and they are quite
widely distributed in action space.  Nonetheless, the stars that lie within
$1.5\kpc$ of the Sun are concentrated into a grid of patches in action space
because only stars with certain frequencies are currently near the Sun.  To
see the patchiness of the distribution in action space it is not necessary to
use the angle-action coordinates of the true Galactic potential. But the
correct potential must be used if statistical measures constructed from the
angle coordinates of stars are to show a characteristic pattern of beats from
which the time at which  the relic was disrupted can be deduced. Hence our results suggest a
two-stage procedure: first a reasonable approximation to the Galactic
potential is used to identify relics through the clustering of their stars'
points around the nodes of a grid in action space. Then once a relic has been
identified, the Galactic potential would be adjusted until the angle-variable
diagnostics showed pronounced beats. This second step would not only pin down
the Galaxy's potential, but also reveal the time at which  relic was
disrupted. 

Growth in the mass of the disc since the satellite fell in would have
significant effects only on Fig.~\ref{fig:stat}: to recover this plot it
would be necessary to model the time dependence of the Galactic potential, so
that the integrals $\int dt\,\bolom$ could be evaluated. We anticipate that
with the help of angle-action coordinates this could be done to sufficient
accuracy, but defer this refinement to a subsequent publication.

We have neglected the deviations of the Galactic potential from axisymmetry.
Could these deviations have a significant impact? \cite{Juric2008} use star
counts in the SDSS survey to show that the Galaxy's thick disc is remarkably
axisymmetric near the Sun. This finding suggests that it is
legitimate to neglect the bar when searching for relics within the
thick disc, such as the Arcturus group. In general, the quadrupole moments of
the bar's gravitational potential will decline rapidly outside the end of the
bar at $R\sim3\kpc$, so stars that are not resonant with the bar will not be
strongly affected by it. The observed axisymmetry of the thick disc suggests
that few if any of its stars are resonantly trapped by the bar, so their
orbits can be safely modelled with an axisymmetric potential. The question
of how the -- phase-dependent -- effects of the bar would impact 
Fig.~\ref{fig:stat} may prove important.

The exciting possibilities discussed here rest on two foundations. One is the
availability of angle-action coordinates for any given potential, and the
other is the availability of full phase space coordinates for significant
samples of stars.  The torus construction technique developed in a series of
paper starting with \cite{McGB90} can provide angle-action coordinates, and
programmes such as the Geneva-Copenhagen, RAVE and Gaia surveys will provide
the phase space coordinates.

Currently the torus technique is restricted to either axisymmetric systems or
two-dimensional non-rotating bars. However, extension to three-dimensional
bars, including bars that are rotating with a constant pattern speed, 
is in principle straightforward and will be attempted soon. 

Clearly when this technique is used to search a real catalogue for relics,
and then to analyse them, difficulties will be encountered that we have
ignored here. Most obviously one will have to contend with errors in the
phase space coordinates of stars (primarily due to errors in distances) and
with the difficulty in picking out overdensities in action space against a
background of Poisson noise from field stars. We are currently applying the
method to $\sim200\,000$ stars from the RAVE survey and hope to report the
results in the near future.

\section*{Acknowledgments}

We are grateful to Walter Dehnen for making his torus code available to us.
We also thank Ben Burnett, John Magorrian and Andy Eyre and the other members 
of the Oxford
dynamics group for critical comments on this work. PJM is supported by a
grant from the Science and Technology Facilities Council.

 \bibliographystyle{mn2e}

\bibliography{refs}

\begin{thebibliography}{32}
\expandafter\ifx\csname natexlab\endcsname\relax\def\natexlab#1{#1}\fi

\bibitem[{{Abadi} {et~al.}(2003){Abadi}, {Navarro}, {Steinmetz}, \&
  {Eke}}]{Abadietal2003}
{Abadi} M.~G., {Navarro} J.~F., {Steinmetz} M., {Eke} V.~R., 2003, \apj, 597,
  21

\bibitem[{{Belokurov} {et~al.}(2006){Belokurov}, {Zucker}, {Evans}, {Gilmore},
  {Vidrih}, {Bramich}, {Newberg}, {Wyse}, {Irwin}, {Fellhauer}, {Hewett},
  {Walton}, {Wilkinson}, {Cole}, {Yanny}, {Rockosi}, {Beers}, {Bell},
  {Brinkmann}, {Ivezi{\'c}}, \& {Lupton}}]{FieldofStreams}
{Belokurov} V., {Zucker} D.~B., {Evans} N.~W., {Gilmore} G., {Vidrih} S.,
  {Bramich} D.~M., {Newberg} H.~J., {Wyse} R.~F.~G., {Irwin} M.~J., {Fellhauer}
  M., {Hewett} P.~C., {Walton} N.~A., {Wilkinson} M.~I., {Cole} N., {Yanny} B.,
  {Rockosi} C.~M., {Beers} T.~C., {Bell} E.~F., {Brinkmann} J., {Ivezi{\'c}}
  {\v Z}., {Lupton} R., 2006, \apjl, 642, L137

\bibitem[{{Binney} \& {Tremaine}(2008)}]{BT08}
{Binney} J., {Tremaine} S., 2008, {Galactic dynamics}. Princeton, NJ, Princeton
  University Press

\bibitem[{{Chereul} {et~al.}(1999){Chereul}, {Cr{\'e}z{\'e}}, \&
  {Bienaym{\'e}}}]{Chereuletal1999}
{Chereul} E., {Cr{\'e}z{\'e}} M., {Bienaym{\'e}} O., 1999, \aaps, 135, 5

\bibitem[{{Dehnen}(1999)}]{Dehnen1999:Bar}
{Dehnen} W., 1999, \apjl, 524, L35

\bibitem[{{Dehnen}(2000)}]{Dehnen2000:OLR}
---, 2000, \aj, 119, 800

\bibitem[{{Dehnen}(2002)}]{Dehnen2002}
---, 2002, Journal of Computational Physics, 179, 27

\bibitem[{{Dehnen} \& {Binney}(1998)}]{DehnenBinney1998}
{Dehnen} W., {Binney} J., 1998, \mnras, 294, 429

\bibitem[{{Eggen}(1971)}]{Eggen1971}
{Eggen} O.~J., 1971, \pasp, 83, 271

\bibitem[{{Famaey} {et~al.}(2005){Famaey}, {Jorissen}, {Luri}, {Mayor}, {Udry},
  {Dejonghe}, \& {Turon}}]{Famaeyetal2005}
{Famaey} B., {Jorissen} A., {Luri} X., {Mayor} M., {Udry} S., {Dejonghe} H.,
  {Turon} C., 2005, \aap, 430, 165

\bibitem[{{Freeman} \& {Bland-Hawthorn}(2002)}]{FreemanBH2002}
{Freeman} K., {Bland-Hawthorn} J., 2002, \araa, 40, 487

\bibitem[{{Fux}(2001)}]{Fux2001}
{Fux} R., 2001, \aap, 373, 511

\bibitem[{{Helmi} \& {de Zeeuw}(2000)}]{HelmideZeeuw2000}
{Helmi} A., {de Zeeuw} P.~T., 2000, \mnras, 319, 657

\bibitem[{{Helmi} {et~al.}(2006){Helmi}, {Navarro}, {Nordstr{\"o}m},
  {Holmberg}, {Abadi}, \& {Steinmetz}}]{Helmietal2006}
{Helmi} A., {Navarro} J.~F., {Nordstr{\"o}m} B., {Holmberg} J., {Abadi} M.~G.,
  {Steinmetz} M., 2006, \mnras, 365, 1309

\bibitem[{{Helmi} \& {White}(1999)}]{HelmiWhite1999}
{Helmi} A., {White} S.~D.~M., 1999, \mnras, 307, 495

\bibitem[{{Helmi} {et~al.}(1999){Helmi}, {White}, {de Zeeuw}, \&
  {Zhao}}]{Helmietal1999}
{Helmi} A., {White} S.~D.~M., {de Zeeuw} P.~T., {Zhao} H., 1999, \nat, 402, 53

\bibitem[{{Helmi} {et~al.}(2003){Helmi}, {White}, \&
  {Springel}}]{Helmietal2003}
{Helmi} A., {White} S.~D.~M., {Springel} V., 2003, \mnras, 339, 834

\bibitem[{{Ibata} {et~al.}(1994){Ibata}, {Gilmore}, \&
  {Irwin}}]{IbataGilmoreIrwin1994}
{Ibata} R.~A., {Gilmore} G., {Irwin} M.~J., 1994, \nat, 370, 194

\bibitem[{{Juri{\'c}} {et~al.}(2008){Juri{\'c}}, {Ivezi{\'c}}, {Brooks},
  {Lupton}, {Schlegel}, {Finkbeiner}, {Padmanabhan}, {Bond}, {Sesar},
  {Rockosi}, {Knapp}, {Gunn}, {Sumi}, {Schneider}, {Barentine}, {Brewington},
  \& {Brinkmann}}]{Juric2008}
{Juri{\'c}} M., {Ivezi{\'c}} {\v Z}., {Brooks} A., {Lupton} R.~H., {Schlegel}
  D., {Finkbeiner} D., {Padmanabhan} N., {Bond} N., {Sesar} B., {Rockosi}
  C.~M., {Knapp} G.~R., {Gunn} J.~E., {Sumi} T., {Schneider} D.~P., {Barentine}
  J.~C., {Brewington} H.~J., {Brinkmann} J., 2008, \apj, 673, 864

\bibitem[{{Kaasalainen} \& {Binney}(1994)}]{KaasB94}
{Kaasalainen} M., {Binney} J., 1994, \mnras, 268, 1033

\bibitem[{{Lynden-Bell} \& {Lynden-Bell}(1995)}]{LyndenBell1995}
{Lynden-Bell} D., {Lynden-Bell} R.~M., 1995, \mnras, 275, 429

\bibitem[{{McGill} \& {Binney}(1990)}]{McGB90}
{McGill} C., {Binney} J., 1990, \mnras, 244, 634

\bibitem[{{Navarro} {et~al.}(2004){Navarro}, {Helmi}, \&
  {Freeman}}]{NavarroHelmiFreeman2004}
{Navarro} J.~F., {Helmi} A., {Freeman} K.~C., 2004, \apjl, 601, L43

\bibitem[{{Nordstr{\"o}m} {et~al.}(2004){Nordstr{\"o}m}, {Mayor}, {Andersen},
  {Holmberg}, {Pont}, {J{\o}rgensen}, {Olsen}, {Udry}, \&
  {Mowlavi}}]{Nordstrometal2004}
{Nordstr{\"o}m} B., {Mayor} M., {Andersen} J., {Holmberg} J., {Pont} F.,
  {J{\o}rgensen} B.~R., {Olsen} E.~H., {Udry} S., {Mowlavi} N., 2004, \aap,
  418, 989

\bibitem[{{Perryman} {et~al.}(2001){Perryman}, {de Boer}, {Gilmore}, {H{\o}g},
  {Lattanzi}, {Lindegren}, {Luri}, {Mignard}, {Pace}, \& {de Zeeuw}}]{GAIA2001}
{Perryman} M.~A.~C., {de Boer} K.~S., {Gilmore} G., {H{\o}g} E., {Lattanzi}
  M.~G., {Lindegren} L., {Luri} X., {Mignard} F., {Pace} O., {de Zeeuw} P.~T.,
  2001, \aap, 369, 339

\bibitem[{{Press} {et~al.}(1986){Press}, {Flannery}, \&
  {Teukolsky}}]{Pressetal1986}
{Press} W.~H., {Flannery} B.~P., {Teukolsky} S.~A., 1986, {Numerical recipes.
  The art of scientific computing}. Cambridge: University Press

\bibitem[{{Sellwood} \& {McGaugh}(2005)}]{SellwoodMcgaugh05}
{Sellwood} J.~A., {McGaugh} S.~S., 2005, \apj, 634, 70

\bibitem[{{Springel} \& {Hernquist}(2003)}]{SpringelHernquist2003}
{Springel} V., {Hernquist} L., 2003, \mnras, 339, 289

\bibitem[{{Steinmetz} {et~al.}(2006){Steinmetz}, {Zwitter}, {Siebert},
  {Watson}, {Freeman}, {Munari}, {Campbell}, {Williams}, {Seabroke}, {Wyse},
  {Parker}, {Bienaym{\'e}}, {Roeser}, {Gibson}, {Gilmore}, {Grebel}, {Helmi},
  {Navarro}, {Burton}, {Cass}, {Dawe}, {Fiegert}, {Hartley}, {Russell},
  {Saunders}, {Enke}, {Bailin}, {Binney}, {Bland-Hawthorn}, {Boeche}, {Dehnen},
  {Eisenstein}, {Evans}, {Fiorucci}, {Fulbright}, {Gerhard}, {Jauregi}, {Kelz},
  {Mijovi{\'c}}, {Minchev}, {Parmentier}, {Pe{\~n}arrubia}, {Quillen}, {Read},
  {Ruchti}, {Scholz}, {Siviero}, {Smith}, {Sordo}, {Veltz}, {Vidrih}, {von
  Berlepsch}, {Boyle}, \& {Schilbach}}]{Steinmetz2006}
{Steinmetz} M., {Zwitter} T., {Siebert} A., {Watson} F.~G., {Freeman} K.~C.,
  {Munari} U., {Campbell} R., {Williams} M., {Seabroke} G.~M., {Wyse} R.~F.~G.,
  {Parker} Q.~A., {Bienaym{\'e}} O., {Roeser} S., {Gibson} B.~K., {Gilmore} G.,
  {Grebel} E.~K., {Helmi} A., {Navarro} J.~F., {Burton} D., {Cass} C.~J.~P.,
  {Dawe} J.~A., {Fiegert} K., {Hartley} M., {Russell} K.~S., {Saunders} W.,
  {Enke} H., {Bailin} J., {Binney} J., {Bland-Hawthorn} J., {Boeche} C.,
  {Dehnen} W., {Eisenstein} D.~J., {Evans} N.~W., {Fiorucci} M., {Fulbright}
  J.~P., {Gerhard} O., {Jauregi} U., {Kelz} A., {Mijovi{\'c}} L., {Minchev} I.,
  {Parmentier} G., {Pe{\~n}arrubia} J., {Quillen} A.~C., {Read} M.~A., {Ruchti}
  G., {Scholz} R.-D., {Siviero} A., {Smith} M.~C., {Sordo} R., {Veltz} L.,
  {Vidrih} S., {von Berlepsch} R., {Boyle} B.~J., {Schilbach} E., 2006, \aj,
  132, 1645

\bibitem[{{Tremaine}(1999)}]{Tremaine99}
{Tremaine} S., 1999, \mnras, 307, 877

\bibitem[{{White} \& {Rees}(1978)}]{WhiteRees1978}
{White} S.~D.~M., {Rees} M.~J., 1978, \mnras, 183, 341

\bibitem[{{Zwitter} {et~al.}(2008){Zwitter}, {Siebert}, {Munari}, {Freeman},
  {Siviero}, {Watson}, {Fulbright}, {Wyse}, {Campbell}, {Seabroke}, {Williams},
  {Steinmetz}, {Bienaym{\'e}}, {Gilmore}, {Grebel}, {Helmi}, {Navarro},
  {Anguiano}, {Boeche}, {Burton}, {Cass}, {Dawe}, {Fiegert}, {Hartley},
  {Russell}, {Veltz}, {Bailin}, {Binney}, {Bland-Hawthorn}, {Brown}, {Dehnen},
  {Evans}, {Re Fiorentin}, {Fiorucci}, {Gerhard}, {Gibson}, {Kelz}, {Kujken},
  {Matijevi{\v c}}, {Minchev}, {Parker}, {Pe{\~n}arrubia}, {Quillen}, {Read},
  {Reid}, {Roeser}, {Ruchti}, {Scholz}, {Smith}, {Sordo}, {Tolstoi},
  {Tomasella}, {Vidrih}, \& {de Boer}}]{Zwitter2008}
{Zwitter} T., {Siebert} A., {Munari} U., {Freeman} K.~C., {Siviero} A.,
  {Watson} F.~G., {Fulbright} J.~P., {Wyse} R.~F.~G., {Campbell} R., {Seabroke}
  G.~M., {Williams} M., {Steinmetz} M., {Bienaym{\'e}} O., {Gilmore} G.,
  {Grebel} E.~K., {Helmi} A., {Navarro} J.~F., {Anguiano} B., {Boeche} C.,
  {Burton} D., {Cass} P., {Dawe} J., {Fiegert} K., {Hartley} M., {Russell} K.,
  {Veltz} L., {Bailin} J., {Binney} J., {Bland-Hawthorn} J., {Brown} A.,
  {Dehnen} W., {Evans} N.~W., {Re Fiorentin} P., {Fiorucci} M., {Gerhard} O.,
  {Gibson} B., {Kelz} A., {Kujken} K., {Matijevi{\v c}} G., {Minchev} I.,
  {Parker} Q.~A., {Pe{\~n}arrubia} J., {Quillen} A., {Read} M.~A., {Reid} W.,
  {Roeser} S., {Ruchti} G., {Scholz} R.-D., {Smith} M.~C., {Sordo} R.,
  {Tolstoi} E., {Tomasella} L., {Vidrih} S., {de Boer} E.~W., 2008, \aj, 136,
  421

\end{thebibliography}
 
\end{document}